# Quantum Correlations and Bell Inequality Violation under Decoherence

Volkan Erol[1,*]

[1] *Computer Engineering Department, Okan University, Istanbul, 34959, Turkey*
*E-mail:* volkan.erol@gmail.com

**Abstract**

Quantum Correlations are studied extensively in quantum information domain. Entanglement Measures and Quantum Discord are good examples of these actively studied correlations. Detection of violation in Bell inequalities is also a widely active area in quantum information theory world. In this work, we revisit the problem of analyzing the behavior of quantum correlations and violation of Bell inequalities in noisy channels. We extend the problem defined in [1] by observing the changes in negativity measure, quantum discord and a modified version of Horodecki measure for violation of Bell inequalities under amplitude damping, phase damping and depolarizing channels. We report different interesting results for each of these correlations and measures. All these correlations and measures decrease under decoherence channels, but some changes are very dramatical comparing to others. We investigate also separability conditions of example studied states.

Keywords

Negativity; Quantum Discord; Violation of Bell inequalities; Decoherence

1. Introduction

Quantum information theory and quantum computing are theoritical basis of quantum computers. Thanks to entanglement, quantum mechanical systems are provisioned to realize many information processing problems faster than classical counterparts. For example, Shor's factorization algorithm, Grover's search algorithm, quantum Fourrier transformation, etc. Entanglement, is the theoretical basis providing the expected speedups. It can be view in bipartite or multipartite forms. In order to quantify entanglement, some measures are defined. On the other hand, a general and accepted criterion, which can measure the amount of entanglement of multilateral systems, has not yet been found.

However, since it must be used in many information processing tasks, the production and processing of multilateral quantum entangled systems is at the top of the hot topics of recent years [2-9]. Much of the work in the basic quantum technologies, such as quantum cryptography, communications, and computers, requires multi-partite entangled systems such as GHZ, W [10,11]. It can be suggested that the quantum entanglement criteria reflects the different properties of the systems. Many recent research has been done in entanglement and its related disciplines like entanglement measures, majorization, quantum fisher information, etc [12-24,32].

Entanglement can be defined as quantum correlations between multiple quantum systems. In this case, the question posed is what does quantum correlation look like and what is different from classical correlation? The discussion on 'quantum' and 'classical' effects is a hot topic. We can define classical correlations in the context of quantum information as those arising from the use of *LOCC*. If we look at a quantum system and can not simulate them classically, we generally have *quantum correlations*. Suppose we have a noisy quantum system and we are working on it on LOCC. In this process we can obtain such a system state that we can do some things we can not achieve with classical correlations, such as violating Bell inequality. In this case, we can obtain these effects by quantum correlations (like quantum discord) in the initial system state that are already present at the source location (even if it is a very noisy system state), not after the LOCC operations. This is the most important point of the entanglement studies.

Until now there has been few studies focused on the relation between quantum discord and Bell inequalities violation [1]. In this work, we extend the problem defined in [1] for two other noisy channels: phase damping (PDC) and depolarizing (DPC). Especially, for these correlations and measures we proved the following:

- Under DPC, the studied state becomes separable very rapidly. Negativity measure for this type states is more robust to ADC and PDC than the DPC.

- Under PDC and DPC quantum discord values decreases more rapidly then the ones under ADC. Quantum discord is more robust to ADC than the other noisy channels for this type of quantum states.

- Under PDC, B (modified measure of Bell inequalities violation) values are equal to Negativity values under PDC. After the value of p=0.2928 none of the states are violating Bell inequalities under ADC and after the value of p=0.1591 none of the states are violating Bell inequalities under DPC.

## 2. Materials and Methods

In this section we make the definitions of our scientific materials and methods. First, we define the studied state. Secondly, we definition one of most actively studied entanglement measure Negativity. In third order, we define quantum discord which is a special quantum correlation. Next, we define violation of Bell inequalities and finally we define the decoherence channels which we study here: amplitude damping, phase damping and depolarizing.

### 2.1. Definitions about Studied State

We consider a two-level quantum system whose computational bases are $|0\rangle$ and $|1\rangle$, and then assume that two observer, Alice and Bob which are sensitive only to mode $|n\rangle_A$ and $|n\rangle_B$, are respectively sharing an entangled initial state

$$|\varphi\rangle_{AB} = \alpha|01\rangle + \sqrt{1-\alpha^2}|01\rangle, \tag{1}$$

where the parameter $\alpha \in (0,1)$.

*2.2. Negativity*

Negativity is a quantitative version of Peres-Horodecki criterion. It is defined for two particle two level general quantum systems as follows [25,26]:

$$N(\rho) = \max\{0, -2\mu_{min}\}, \tag{2}$$

Here $\mu_{min}$ value is the minimum eigenvalue of $\rho$'s partial transpose. $\rho$ is the density matrix of the quantum state. Negativity, which is defined by the equation above is a value between 0 and 1 like Concurrence. Similarly like for concurrence, 1 means maximal entanglement. 0 means that the state is a separable state.

Vidal and Werner shown that Negativity is a monotone function for entanglement [26].

*2.3 Quantum Discord*

The definitions used in this subsection are well reviewed and organized in a review by Streltsov [33]. Quantum discord is considered as the first measure of quantum correlations beyond entanglement [27]. The definition of quantum discord is made on the basis of mutual information between two random variables $X$ and $Y$ and it is possible this definition with the following equations

$$I(X:Y) = H(X) + H(Y) - H(X,Y), \tag{3}$$

$$J(X:Y) = H(X) - H(X|Y). \tag{4}$$

Here, $H(X) = -\sum_x p_x \log_2 p_x$ is the random variable $X$'s Shannon entropy where $p_x$ is the probability when $X$ takes the value $x$. $H(X,Y)$ is the joint entropy of both variables $X$ and $Y$. The conditional entropy $H(X/Y)$ can be written as

$$H(X|Y) = \sum_y p_y H(X|y), \tag{5}$$

where $p_y$ is the probability that the random variable $Y$ takes the value $y$, and $H(X/y)$ is the entropy of the variable $X$ conditioned on the variable $Y$ taking the value $y$: $H(X|y) = \sum_x p_{x|y} \log_2 p_{x|y}$, and $p_{x/y}$ is the probability of x given y.

The following equality of I and J is coming from Bayes rule $p_{x/y} = p_{xy}/p_y$, which can be used to show that $H(X/Y)=H(X,Y)-H(Y)$. However, I and J are no longer equal if quantum theory is applied [27, 33]. More precisely, for a quantum state $\rho^{AB}$ the mutual information between A and B can be defined as

$$I(\rho^{AB}) = S(\rho^A) + S(\rho^B) - S(\rho^{AB}) \tag{6}$$

with the von Neumann entropy S, and the reduced density operators $\rho^A = Tr_B[\rho^{AB}]$ and $\rho^B = Tr_A[\rho^{AB}]$. This expression is the generalization of the classical mutual information $I(X:Y)$ to the quantum theory.

On the other hand, the generalization of $J(X:Y)$ is not completely obvious. The following way to generalize $J$ to the quantum theory was proposed [27]: for a bipartite quantum state $\rho^{AB}$ the conditional entropy of A conditioned on a measurement on B was defined as:

$$S(A|\{\Pi_i^B\}) = \sum_i p \, S(\rho_i^A), \tag{7}$$

where $\{\Pi_i^B\}$ are measurement operators corresponding to a von Neumann measurement on the subsystem B, i.e, orthogonal projectors with rank one. The probability $p_i$ for obtaining the outcome $i$ can be given by $p_i = Tr[\Pi_i^B \rho^{AB}]$, and the corresponding post-measurement state of the subsystem A is defined by $\rho_i^A = Tr_B[\Pi_i^B \rho^{AB}]/p_i$. The equality $J$ can now be extended to quantum states as follows [27]:

$$J(\rho^{AB})_{\{\Pi_i^B\}} = S(\rho)^A - S(A|\{\Pi_i^B\}), \tag{8}$$

where the index $\{\Pi_i^B\}$ clarifies that the value depends on the choice of the measurement operators $\Pi_i^B$. The quantity $J$ represents the amount of information gained about the subsystem A by measuring the subsystem B [27].

Quantum discord is defined as the difference of these two inequivalent expressions for the mutual information, minimized over all von Neumann measurements:

$$\delta^{B|A}(\rho^{AB}) = min_{\{\Pi_i^B\}}\left[I(\rho^{AB}) - J(\rho^{AB})_{\{\Pi_i^B\}}\right], \tag{9}$$

where the minimum over all von Neumann measurements is taken in order to have a measurement-independent expression [27, 33]. As was also shown in [27], quantum discord is nonnegative, and is equal to zero on quantum-classical states only. These are states of the form $\rho = \sum_i p_i \, \rho_i^A \otimes |i\rangle\langle i|^B$.

*2.4 Violation of Bell inequality: Measure of CHSH violation*

The CHSH inequality for a two-qubit state $\rho \equiv \rho_{AB}$ can be written as [29,30]:

$$|Tr(\rho B_{CHSH})| \leq 2 \tag{10}$$

in terms of the CHSH operator

$$B_{CHSH} = \vec{a}.\vec{\sigma} \otimes (\vec{b} + \vec{b}').\vec{\sigma} + \vec{a}'.\vec{\sigma} \otimes (\vec{b} - \vec{b}').\vec{\sigma} \tag{11}$$

where $\vec{a}$, $\vec{a}'$ and $\vec{b}$, $\vec{b}'$ are unit vectors describing the measurements on side A (Alice) and B (Bob), respectively. A shown by Horodecki et al. [29,30] by optimizing the vectors $\vec{a}$, $\vec{a}'$, $\vec{b}$, $\vec{b}'$, the maximum possible average value of the Bell operator for the state $\rho$ is given by

$$\max_{B_{CHSH}} |Tr(\rho B_{CHSH})| = 2\sqrt{M(\rho)} \tag{12}$$

where $M(\rho) = \max_{j<k}\{h_j + h_k\} \leq 2$, and $h_j$ ($j=1,2,3$) are the eigenvalues of the matrix $U = T^T T$ constructed from the correlation matrix $T$ and its transpose $T^T$. [28]

In other words, $M(\rho) = \tau_1 + \tau_2$ where $\tau_1$ and $\tau_2$ are maximum eigenvalues of symmetric $U$ matrix.

It can be showed that

$$M(\rho) = \tau_1 + \tau_2 = \max\{c_1^2 + c_2^2, c_1^2 + c_3^2, c_2^2 + c_3^2\} \tag{13}$$

Now our question become calculating $\tau_1 + \tau_2$ for any random states / density matrix.

In order to quantify the violation of CHSH inequality one can use $M(\rho)$ or equivalently

$$B(\rho) = \sqrt{\max[0, M(\rho) - 1]} \tag{14}$$

where $B=0$ means CHSH inequality is not violated and $B=1$ means is maximally violated. [28]

*2.5 Decoherence Channels*

In this section, we give the definitions of the three decoherence channels (ADC, PDC and DPC). In general, the decoherence channels are given in the Kraus representation

$$\varepsilon(\rho) = \sum_\mu E_\mu \rho E_\mu^\dagger, \tag{15}$$

where $E_\mu$ are the Kraus operators that satisfy

$$\sum_\mu E_\mu^\dagger E_\mu = 1, \tag{16}$$

where 1 is an identity matrix. In the following subsections, we give the definitions about the three channels consecutively [31]

### 2.5.1 Amplitude Damping Channel (ADC)

The ADC describes the dissipation process. For a single qubit, the Kraus operators of the ADC are

$$E_0 = \sqrt{s}|0\rangle\langle 0| + |1\rangle\langle 1|, \quad E_1 = \sqrt{p}|1\rangle\langle 0|, \tag{17}$$

where the decoherence strength $p = 1 - s$, represents the probability of decay from the upper level $|0\rangle$ to the lower level $|1\rangle$, with $s = \exp\left(-\frac{\gamma_1 t}{2}\right)$, and $\gamma_1$ is the damping rate [31].

### 2.5.2 Phase Damping Channel (PDC)

The Kraus operators for the PDC are given by

$$E_0 = \sqrt{s}\mathbf{1}, \quad E_1 = \sqrt{p}|0\rangle\langle 0|, \quad E_2 = \sqrt{p}|1\rangle\langle 1|. \tag{18}$$

The PDC is a prototype model of dephasing or pure decoherence. The decoherence strength is comparable with a concrete dephasing model by replacing p with $1 - \exp(-\gamma_2 t/2)$, where $\gamma_2$ is associated with the $T_2 = 1/\gamma_2$ relaxation in spin resonance [31].

### 2.5.3 Depolarizing Channel (DPC)

The Kraus operators of the DPC are given by

$$E_0 = \sqrt{1-p'}\mathbf{1}, \quad E_1 = \sqrt{\frac{p'}{3}}\sigma_x, \quad E_2 = \sqrt{\frac{p'}{3}}\sigma_y, \quad E_3 = \sqrt{\frac{p'}{3}}\sigma_z. \tag{19}$$

where $p' = 3p/4$. For the DPC, the spin is depolarized to the maximally mixed state $1/2$ with probability p or is unchanged with probability $s = 1 - p$.

## 3. Results

In this section we give the numerical results that we achieved during our study. This section is organized as follows: First we give Negativity values (a good example of Entanglement measures) changes in ADC, PDC and DPC decoherence channels. Secondly, we show changes in Quantum Discord (correlations beyond entanglement) under mentioned decoherence channels. And finally, we give the changes in B which is a modified Horodecki measure for Violation of Bell inequalities under the same noisy channels.

*3.1. Negativity under Decoherence Channels*

In this subsection we show changes in Negativity values under three noisy channels respectively. The changes in Figure 1 and 2 which are values under APC and PDC are quite similar. But in Figure 3, under DPC changes in Negativity values are different than the others. Especially when the value p increase, the studied state becomes separable.

For example, under DPC when p = 0.2, the state is separable for values $\alpha < 0.15$ and $\alpha > \sim 0.98$. When p=0.3, the state is separable for values $\alpha < 0.27$ and $\alpha > \sim 0.96$. when p becomes 0.4 the changes are more dramatical. The state is separable for values $\alpha < \sim 0.5$ and $\alpha > \sim 0.85$. For larger p values all states are separable.

We can view that under DPC, the studied state becomes separable very rapidly. Negativity measure for this type states is more robust to ADC and PDC than the DPC.

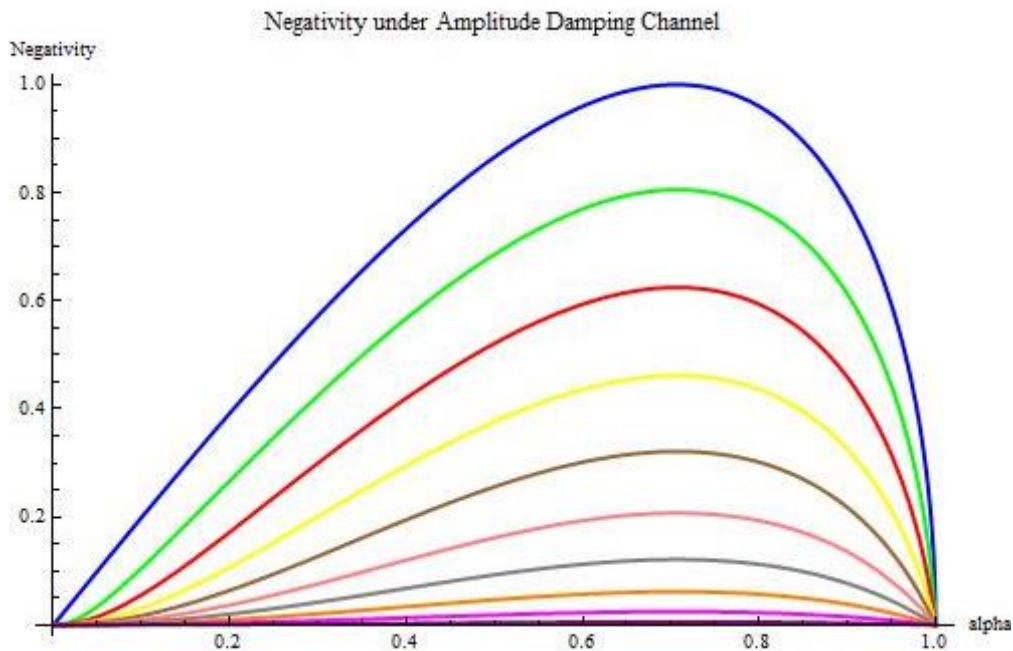

**Figure 1.** Changes in Negativity under the amplitude damping channel (blue: without noise, green: p=0.1, red: p=0.2, yellow: p=0.3, brown: p=0.4, pink: p=0.5, gray: p=0.6, orange: p=0.7, magenta: p=0.8, purple: p=0.9)

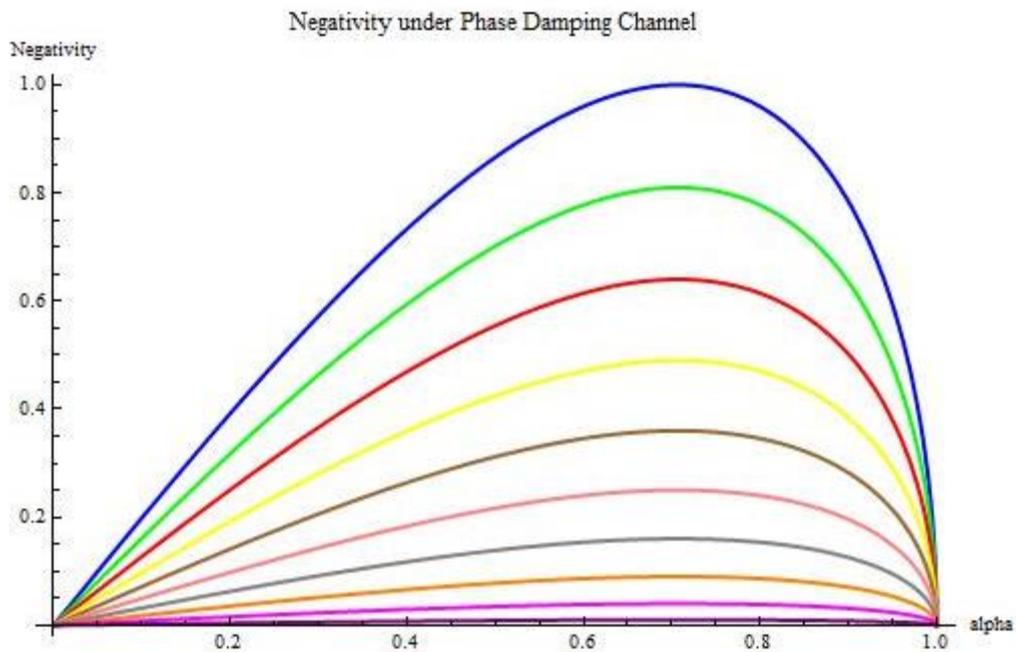

**Figure 2.** Changes in Negativity under the phase damping channel (blue: without noise, green: p=0.1, red: p=0.2, yellow: p=0.3, brown: p=0.4, pink: p=0.5, gray: p=0.6, orange: p=0.7, magenta: p=0.8, purple: p=0.9)

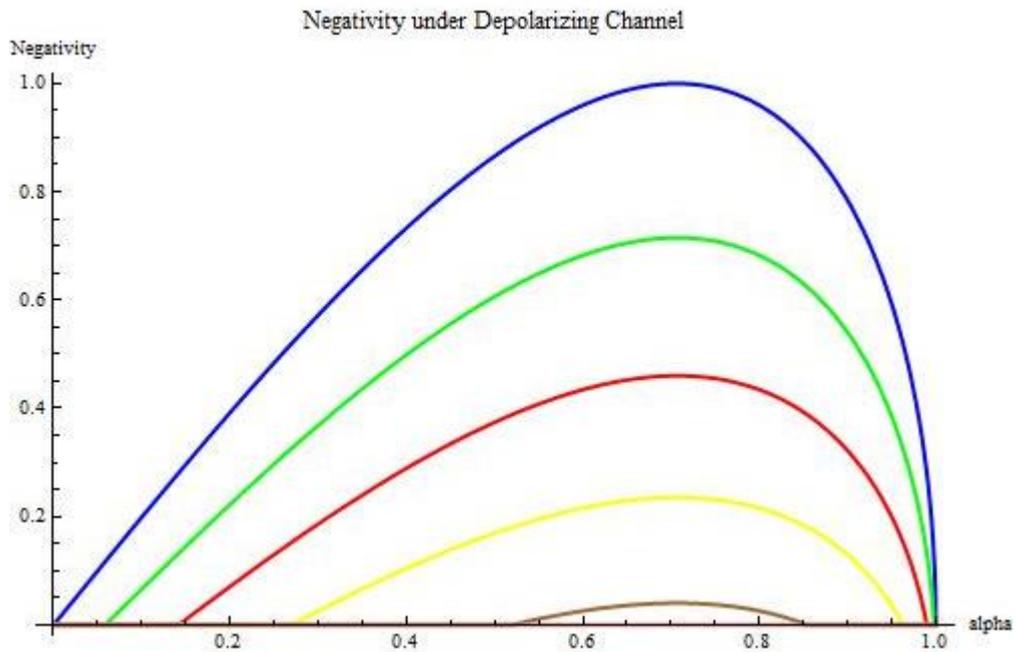

**Figure 3.** Changes in Negativity under the depolarizing channel (blue: without noise, green: p=0.1, red: p=0.2, yellow: p=0.3, brown: p=0.4)

*3.2. Quantum Discord under Decoherence Channels*

In this subsection we show changes in Quantum Discord values under three noisy channels respectively. The changes in Figure 5 and 6 which are values under PDC and DPC are quite similar. In Figure 4 we can see that value changes under ADC quite different. Under PDC and DPC quantum discord values decreases more rapidly then the ones under ADC. Quantum discord is more robust to ADC than the other noisy channels for this type of quantum states.

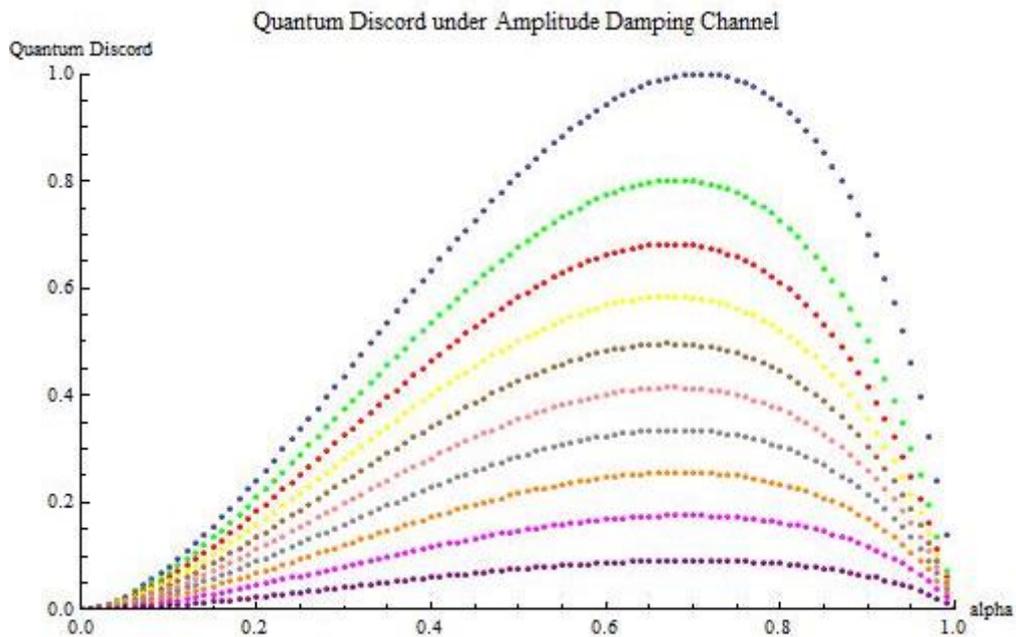

**Figure 4.** Changes in Quantum Discord under the amplitude damping channel (blue: without noise, green: p=0.1, red: p=0.2, yellow: p=0.3, brown: p=0.4, pink: p=0.5, gray: p=0.6, orange: p=0.7, magenta: p=0.8, purple: p=0.9)

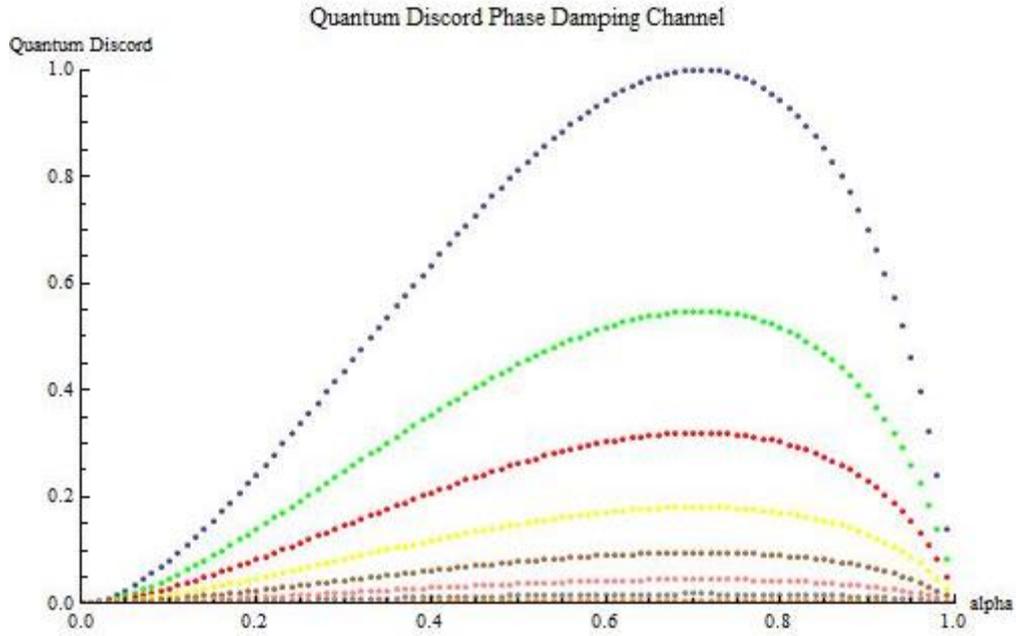

**Figure 5.** Changes in Quantum Discord under the phase damping channel (blue: without noise, green: p=0.1, red: p=0.2, yellow: p=0.3, brown: p=0.4, pink: p=0.5, gray: p=0.6, orange: p=0.7, magenta: p=0.8, purple: p=0.9)

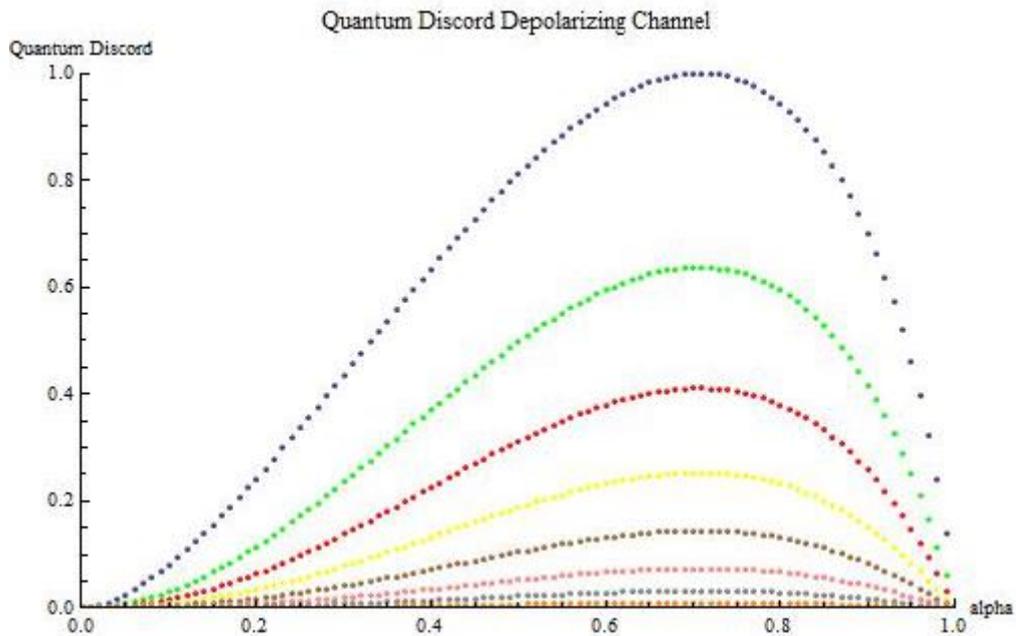

**Figure 6.** Changes in Quantum Discord under the depolarizing channel (blue: without noise, green: p=0.1, red: p=0.2, yellow: p=0.3, brown: p=0.4, pink: p=0.5, gray: p=0.6, orange: p=0.7, magenta: p=0.8, purple: p=0.9)

*3.3. Violation of Bell inequalities (B) under Decoherence Channels*

In this subsection, we give the changes in B which is a modified Horodecki measure for Violation of Bell inequalities under the same noisy channels. In Figure 7, our finding are quite interesting. After the value of p=0.2928 none of the states are violating Bell inequalities under ADC. The changes in B can be viewed in the 3D plot shown in Figure 8.

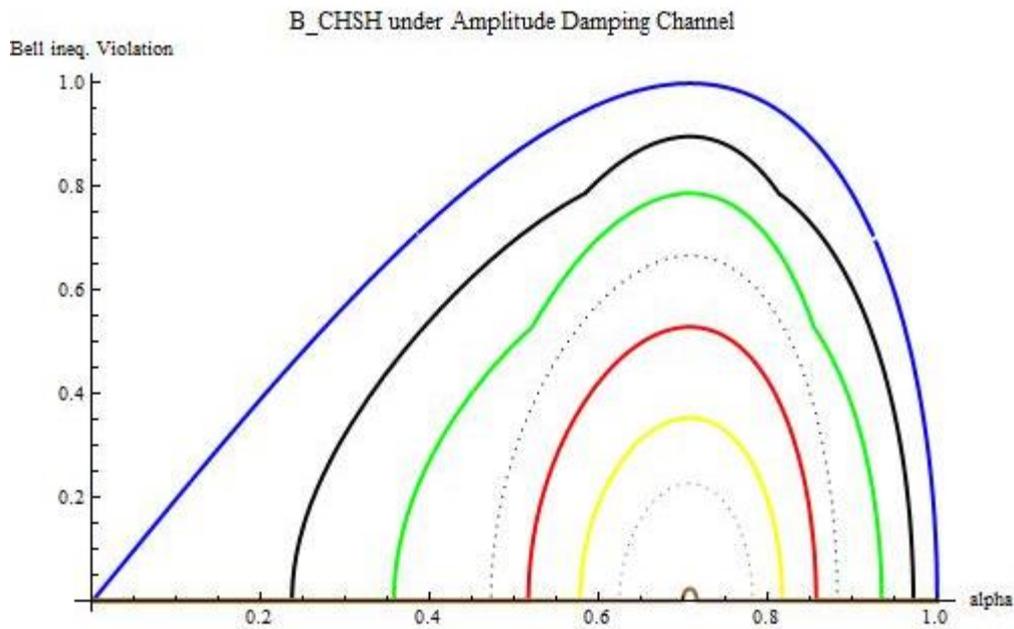

**Figure 7.** Changes in $B_{CHSH}$ under the amplitude damping channel (blue: without noise, black: p=0.05, green: p=0.1, black dotted: p=0.15, red: p=0.2, yellow: p=0.25, brown dotted: p=0.275, brown: p=0.2927)

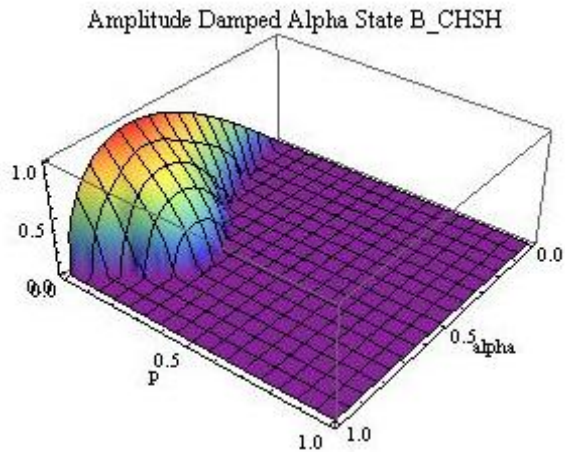

**Figure 8.** Changes in B$_{CHSH}$ under the amplitude damping channel shown as a function of alpha and p.

Under PDC, B values are equal to Negativity values under PDC. This is also a really interesting result achieved in this study. The values can be viewed in Figure 9.

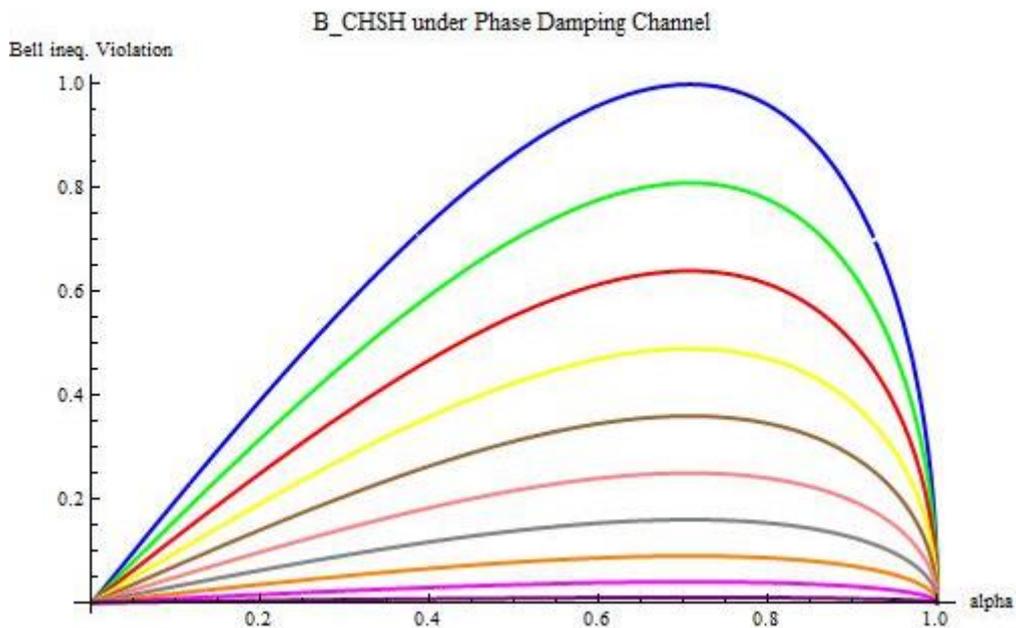

**Figure 9.** Changes in B$_{CHSH}$ under the phase damping channel (blue: without noise, green: p=0.1, red: p=0.2, yellow: p=0.3, brown: p=0.4, pink: p=0.5, gray: p=0.6, orange: p=0.7, magenta: p=0.8, purple: p=0.9)

In Figure 10, our findings are also quite interesting. After the value of p=0.1591 none of the states are violating Bell inequalities under DPC. The changes in B can be viewed in the 3D plot shown in Figure 11.

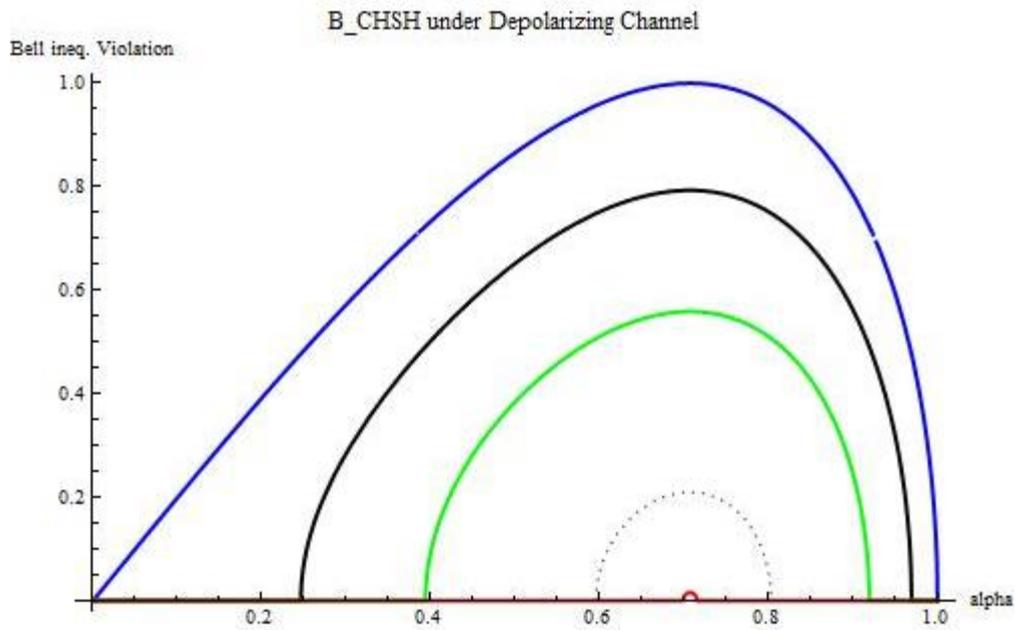

**Figure 10.** Changes in B$_{CHSH}$ under the phase damping channel (blue: without noise, black: p=0.05, green: p=0.1, dotted black: p=0.15, red: p=0.15905)

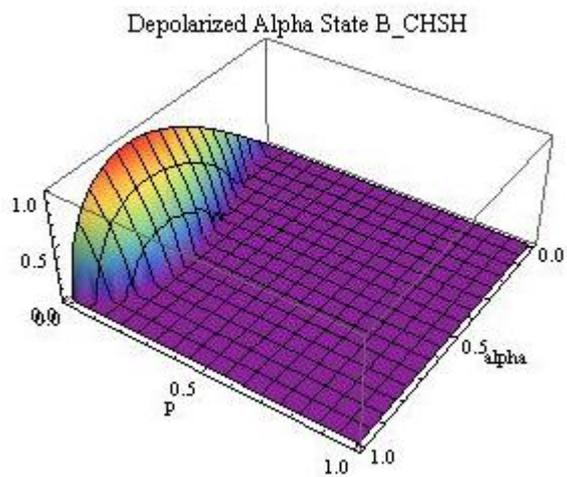

**Figure 11.** Changes in B$_{CHSH}$ under the depolarizing channel shown as a function of alpha and p.

We show in the Figure 12 all the value changes.

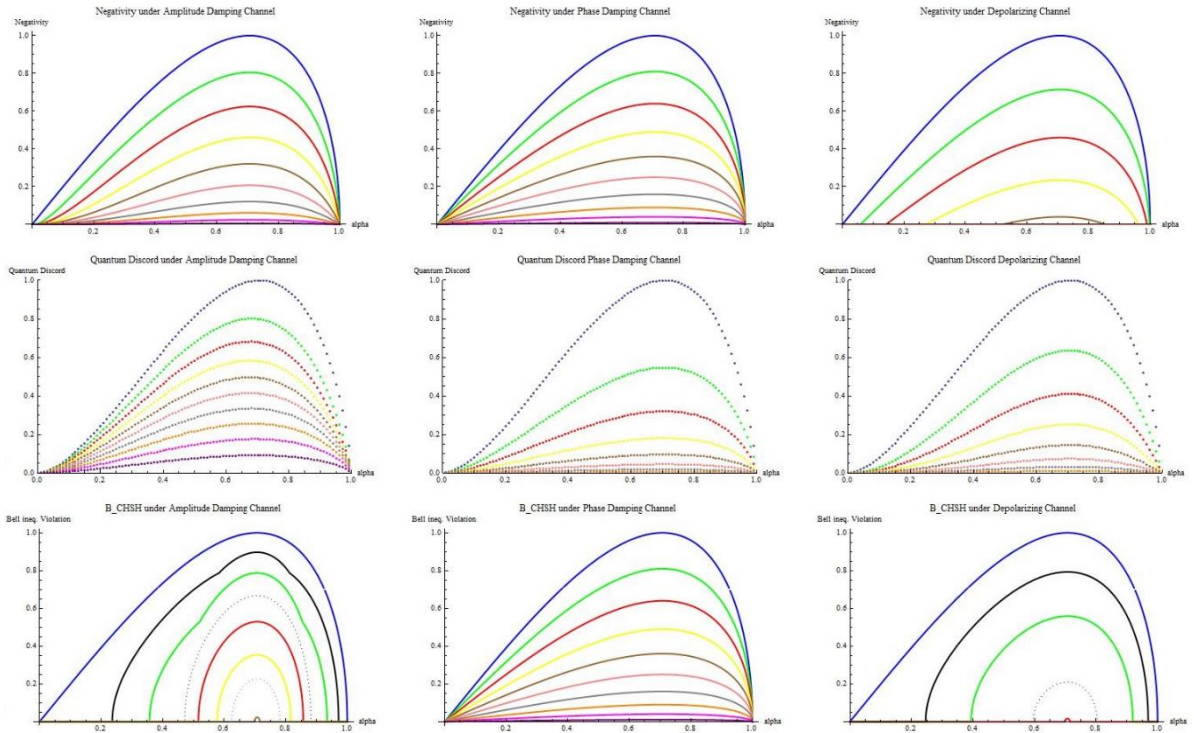

**Figure 12.** Quantum Correlations and Violation of Bell inequality under decoherence channels.

In Figure 13, we show in 3D plot the changes in Negativity and B values as functions of p and α.

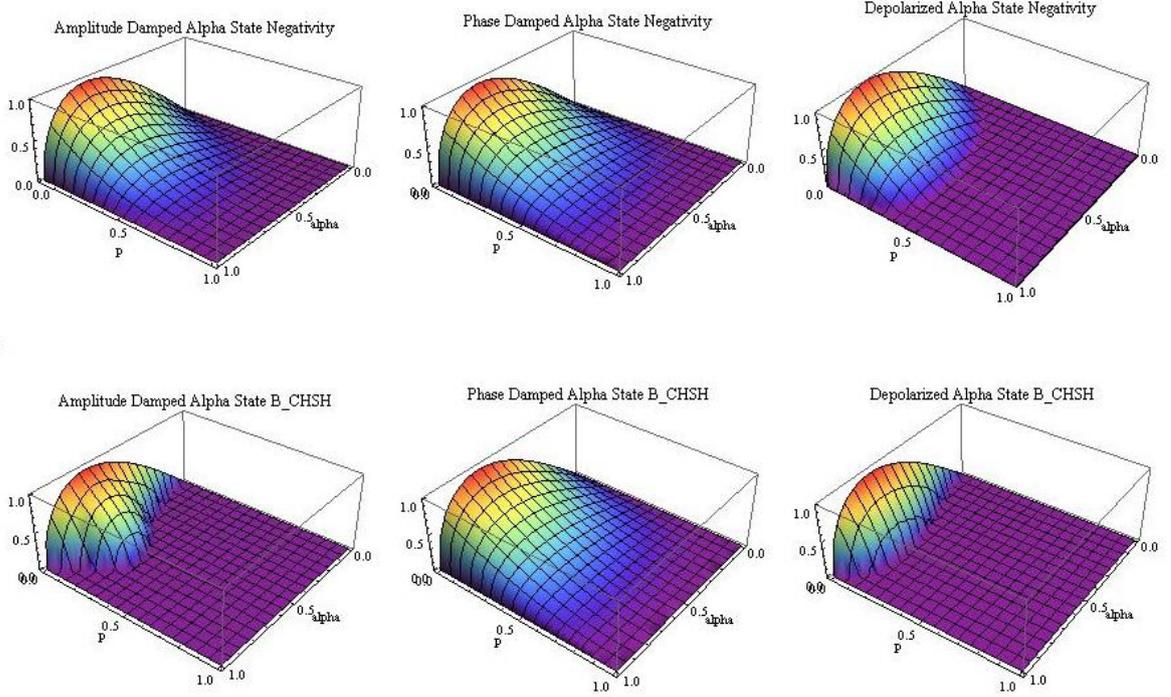

**Figure 13.** Changes in Negativity and B$_{CHSH}$ under the decoherence channels shown as functions of alpha and p.

**4. Conclusions**

In this work, we extend the problem defined in [1] for two other noisy channels: phase damping (PDC) and depolarizing (DPC). We can summarize the following observations achieved during this study:

For the Negativity case, especially when the value decoherence strength p increase, the studied state becomes separable. Under DPC, the studied state becomes separable more rapidly and Negativity measure for this type states is more robust to ADC and PDC than the DPC. For example, under DPC when p = 0.2, the state is separable for values α< 0.15 and α > ~0.98. When p=0.3, the state is separable for values α< 0.27 and α > ~0.96. when p becomes 0.4 the changes are more dramatical. The state is separable for values α< ~0.5 and α > ~0.85. For larger p values all states are separable.

For the Quantum Discord case; under PDC and DPC quantum discord values decreases more rapidly then the ones under ADC. Quantum discord is more robust to ADC than the other noisy channels for this type of quantum states.

For the Violation in Bell inequalities case; under PDC, B (modified measure of Bell inequalities violation) values are equal to Negativity values under PDC. After the value of p=0.2928 none of the states are violating Bell inequalities under ADC and after the value of p=0.1591 none of the states are violating Bell inequalities under DPC. The decrease in DPC is more dramarical than the ones under other decoherence channels.

# References

1. W. Ma, S. Xu, J. Shi and L. Ye, Quantum correlation versus Bell-inequality violation under the amplitude damping channel, Phys. Lett. A, 379, 2802 (2015).

2. M.A. Nielsen and I.L. Chuang, Quantum Computation and Quantum Information, Cambridge, UK: Cambridge Univ. Press, 2000.

3. A. Zeilinger, M. A. Horne, H. Weinfurter, and M. Zukowski, Three-Particle Entanglements from Two Entangled Pairs, Phys. Rev. Lett., vol. 78, no. 3031, 1997.

4. T. Tashima, S. K. Ozdemir, T. Yamamoto, M. Koashi, and N. Imoto, Elementary optical gate for expanding an entanglement web, Phys. Rev. A, vol. 77, no. 030302, 2008.

5. T. Tashima, S. K. Ozdemir, T. Yamamoto, M. Koashi, and N. Imoto, Local expansion of photonic W state using a polarization-dependent beamsplitter, New J. Phys. A, vol. 11, no. 023024, 2009.

6. T. Tashima, T. Wakatsuki, S. K. Ozdemir, T. Yamamoto, M. Koashi, and N. Imoto, Local Transformation of Two Einstein-Podolsky-Rosen Photon Pairs into a Three-Photon W State, Phys. Rev. Lett., vol. 102, no. 130502, 2009.

7. S. Bugu, C. Yesilyurt and F. Ozaydin, Enhancing the W-state quantum-network-fusion process with a single Fredkin gate, Phys. Rev. A, vol. 87, no. 032331, 2013.

8. C. Yesilyurt, S. Bugu and F. Ozaydin, An Optical Gate for Simultaneous Fusion of Four Photonic W or Bell States, Quant. Info. Proc., vol. 12, no. 2965, 2013.

9. F.Ozaydin et al., Fusing multiple W states simultaneously with a Fredkin gate, Phys. Rev. A, vol. 89, no. 042311, 2014.

10. Toth et al., Spin squeezing and entanglement, Phys. Rev. A, vol. 79, no. 042334, 2009.

11. M.Tsang, J.H.Shapiro, S.Lloyd, Quantum Optical Temporal Phase Estimation by Homodyne Phase-Locked Loops, IEEE Conference on Lasers and Electro-Optics (CLEO) 2009.

12. V. Erol, F. Ozaydin, A. A. Altintas, Analysis of entanglement measures and locc maximized quantum fisher information of general two qubit systems, Sci. Rep. 4, 5422, 2014.

13. F. Ozaydin, A. A. Altintas, C. Yesilyurt, S. Bugu, V. Erol, Quantum Fisher Information of Bipartitions of W States, Acta Physica Polonica A 127, 1233-1235, 2015.

14. V. Erol, S. Bugu, F. Ozaydin, A. A. Altintas, An analysis of concurrence entanglement measure and quantum fisher information of quantum communication networks of two-qubits, Proceedings of IEEE 22nd Signal Processing and Communications Applications Conference (SIU2014), pp. 317-320, 2014.

15. V. Erol, A comparative study of concurrence and negativity of general three-level quantum systems of two particles, AIP Conf. Proc. 1653 (020037), 2015.

16. V. Erol, F. Ozaydin, A. A. Altintas, Analysis of Negativity and Relative Entropy of Entanglement measures for qubit-qutrit Quantum Communication systems, Proceedings of IEEE 23rd Signal Processing and Communications Applications Conference (SIU2015), pp. 116-119, 2014.

17. V. Erol, Detecting Violation of Bell Inequalities using LOCC Maximized Quantum Fisher Information and Entanglement Measures, Preprints 2017, 2017030223 (doi: 10.20944/preprints201703.0223.v1), 2017.

18. V. Erol, Analysis of Negativity and Relative Entropy of Entanglement Measures for Two Qutrit Quantum Communication Systems, Preprints 2017, 2017030217 (doi: 10.20944/preprints201703.0217.v1), 2017.

19. V. Erol, The relation between majorization theory and quantum information from entanglement monotones perspective, AIP Conf. Proc. 1727 (020007), 2016.

20. V. Erol, A Proposal for Quantum Fisher Information Optimization and its Relation with Entanglement Measures, PhD Thesis, Okan University, Institute of Science, 2015.

21. V. Erol, Quantum Fisher Information: Theory and Applications. Preprints 2017, 2017040134 (doi: 10.20944/preprints201704.0134.v1), 2017.

22. V. Erol, Entanglement Monotones and Measures: An Overview. Preprints 2017, 2017040098 (doi: 10.20944/preprints201704.0098.v3), 2017.

23. V. Erol, Quantum Fisher Information of W and GHZ State Superposition under Decoherence. Preprints 2017, 2017040147 (doi: 10.20944/preprints201704.0147.v1), 2017.

24. V. Erol, Quantum Fisher Information of Decohered W and GHZ Superposition States with Arbitrary Relative Phase. Preprints 2017, 2017040182 (doi: 10.20944/preprints201704.0182.v1), 2017.

25. R. F. Werner, Quantum states with Einstein-Podolsky-Rosen correlations admitting a hidden-variable model, Phys. Rev. A 40, 4277 (1989).

26. G. Vidal and R. F. Werner, Computable measure of entanglement, Phys. Rev. A 65, 032314 (2002).

27. H. Ollivier and W. H. Zurek, Quantum Discord: A Measure of the Quantumness of Correlations, Phys. Rev. Lett. 88, 017901 (2001).

28. K. Bartkiewicz, B. Horst, K. Lemr, and Adam Miranowicz, Entanglement estimation from Bell inequality violation, Phys. Rev. A 88, 052105 (2013).

29. R. Horodecki, P. Horodecki, and M. Horodecki, Violating Bell inequality by mixed spin-12 states: necessary and sufficient condition, Phys. Lett. A 200, 340 (1995).

30. R. Horodecki, Two-spin-12 mixtures and Bell's inequalities, Phys. Lett. A 210, 223 (1996).

31. J. Ma, Y. Huang, X. Wang, and C. P. Sun, Quantum Fisher information of the Greenberger-Horne-Zeilinger state in decoherence channel, Phys. Rev. A 84, 022302 (2011).

32. A. A. Altintas, Quantum Fisher Information of an open and noisy system in the steady state, Annals of Physics, 362, pp: 192-198, (2016).

33. A. Streltsov, Quantum Discord and its Role in Quantum Information Theory, SpringerBriefs in Physics (2015).